\documentclass[twocolumn,rmp,aps,showpacs]{revtex4}

\usepackage{dcolumn,graphicx,amsmath,amssymb,pxfonts}

\begin{document}

\title{Prisoners' dilemma in real-world acquaintance networks:\\
  Spikes and quasi-equilibria induced by the interplay between
  structure and dynamics
  }

\author{Petter Holme}
\affiliation{Department of Physics, Ume{\aa} University, 901 87
  Ume{\aa}, Sweden}
\author{Ala Trusina}
\affiliation{Department of Physics, Ume{\aa} University, 901 87
  Ume{\aa}, Sweden}
\affiliation{NORDITA, Blegdamsvej 17, 2100 Copenhagen, Denmark}
\author{Beom Jun \surname{Kim}}
\affiliation{Department of Molecular Science
  and Technology, Ajou University, Suwon 442-749, Korea}
\author{Petter Minnhagen}
\affiliation{NORDITA, Blegdamsvej 17, 2100 Copenhagen, Denmark}
\affiliation{Department of Physics, Ume{\aa} University, 901 87
  Ume{\aa}, Sweden}

\begin{abstract}
  We study Nowak and May's spatial prisoners' dilemma game driven by
  mutations (random choices of suboptimal strategies) on empirical
  social networks. The time evolution of the cooperation level is
  highly complex containing spikes and steps between quasi-stable
  levels. A statistical characterization of the quasi-stable states and
  a study of the mechanisms behind the steps are given. We argue that
  the crucial structural ingredients causing the observed behavior is
  an inhomogeneous degree distribution and that the connections within
  vertices of highest degree are rather sparse. Based on these
  observations we construct model networks with a similarly complex
  time-evolution of the cooperation level.
\end{abstract}


\maketitle

\section{Introduction}
The prisoners' dilemma (PD) game is a powerful metaphor for the situation
where mutual trust and cooperation is beneficial in a long perspective but
egoism and guile can produce big short-term profit. Applications range
from evolution of RNA viruses~\cite{rna}, through westernization of
central Africa~\cite{africa}, to the social situation of soldiers in
trench warfare~\cite{axe:evo} and NASCAR drivers~\cite{nascar}. One of the major
achievements has been to establish the condition for cooperative
behavior when two players meet repeatedly~\cite{axe:evo}. Another
direction has been to investigate the criteria for cooperation to be
stable in social space. In this approach the spatial PD game by Nowak and
May~\cite{nowakmay} has been the basic model; a model where there are
a number $N$ of players interacting only with players
in its immediate surrounding. Traditionally,
spatial games have been studied on regular lattices; but real social
networks are very complexly organized~\cite{wf}---being random to some
extent, but also having structure reflecting the social forces. As a
step in this direction people have studied the spatial PD on
more realistic model networks~\cite{abramson,our:pd}. In this Communication we
pursue this idea to the end and let the players' encounters follow the
ties of empirical social networks. The outcome is a very complex behavior
arising from the interplay between the PD dynamics and the underlying
network structure.

\begin{figure}
  \centering{\resizebox*{8.5cm}{!}{\includegraphics{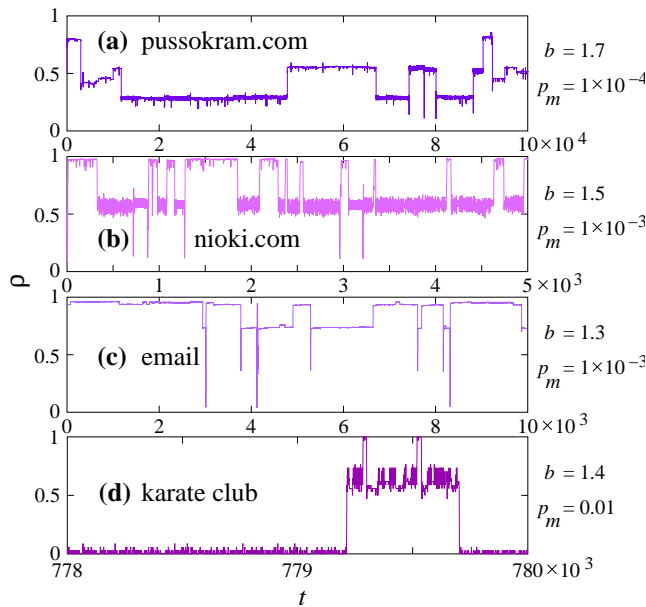}}}
  \caption{Time evolution of the cooperator density $\rho$ for
    different networks and values of the temptation $b$ and mutation
  rate $p_m$ parameters. All these time sequences are typical for the
  given set of parameter values except the karate club network that
  spends most of the time in the all-$D$ state $\rho\approx 0$.
  }
  \label{fig:cd}
\end{figure}

\begin{table}
  \caption{Statistics of the networks: The number of vertices $N$, the
    number of edges $M$, the average degree in the maximal subgraph of
    the ten vertices of highest degree $k_{10}$, and the corresponding
    expectation value for random graphs of the same degree sequence
    $\bar{k}_{10}$. }
\label{tab:nwk}
\begin{ruledtabular}
\begin{tabular}{rr|rr|rr}
  network & Ref.\ &$N$ & $M$ & $k_{10}$ & $\bar{k}_{10}$\\\hline 
  pussokram.com & \cite{pok} & $29{\,}341$ & $115{\,}684$ & $1.6$ & $6.0(1)$\\
  nioki.com & \cite{nioki} & $50{\,}259$ & $239{\,}452$ & $1.2$ & $1.9(1)$\\
  emails & \cite{bornholdt:email} & $40{\,}346$ & $58{\,}224$ & $1.8$
  & $4.9(1)$\\ 
  karate club & \cite{zachary} & $34$ & $78$ & & \\
\end{tabular}
\end{ruledtabular}
\end{table}

We represent the underlying networks as graphs $G=(V,E)$ where $V$ is
the set of $N$ players (or vertices) and $E$ is the set of $M$ ties
(or edges---unordered pairs of vertices) between them. The networks we
use are mostly acquired from online interaction---through contacts
within two
Internet communites (pussokram.com and nioki.com) and through email exchange. (A summary of the used
networks can be seen in Table~\ref{tab:nwk}.) Even if the structure of
online interaction network differs from regular acquaintance
networks, we believe that our results will hold for a quite large
class of social networks. As a small test network we also use an
acquaintance data constructed from a field survey (the ``karate club''
network).

In Nowak and May's spatial PD game~\cite{nowakmay} each
player, at each time step, adopts one of two strategies: cooperate $C$
or defect $D$. To catch the dilemma one lets an encounter between two
cooperators result in unity gain for both players, whereas a $D$-$D$
encounter gives zero gain for both. However, if a cooperator
meets a defector the cooperator scores zero and the defector scores
$b\in (1,2)$ ($b$ is called `temptation'). The gain $g(v)$ of a player
$v$ is the sum of the gains from the encounters with its
neighbors.  If a neighbor
scores higher than a player, the player follows the high scoring
neighbor. But, to drive the system (and model occasional irrational
moves) a player chooses the  opposite strategy with a probability
$p_m$. We use synchronous updating of the players, i.e.\ one time step
of the algorithm consists of one sweep (over all the players) to
calculate the individual gains, and one sweep to update the strategies.

\begin{figure}
  \centering{\resizebox*{7.4cm}{!}{\includegraphics{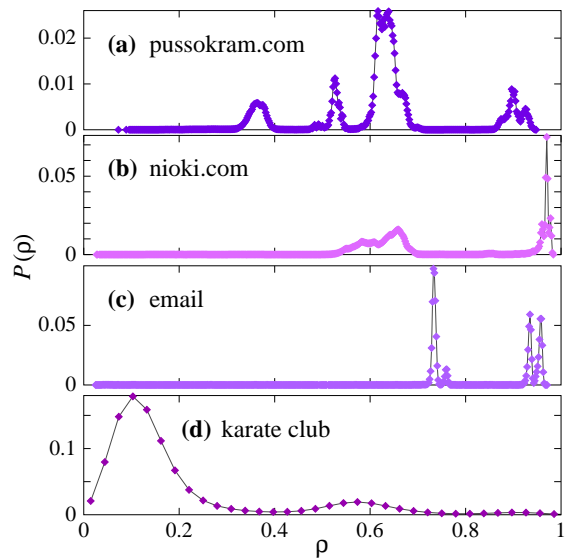}}}
  \caption{
    Histograms (over $10^8$ time steps) of the cooperator density. The
  temptation is $1.35$ for both sub-figures, the mutation rate is
  $0.001$. Lines are guides for the eyes.
  }
  \label{fig:ph}
\end{figure}

\section{Evolution of the cooperator level}

Our key quantity is the cooperator density $\rho$---the fraction of
players adopting the strategy $C$. In regular networks the
cooperator level is characterized by random fluctuations and
high-frequency oscillations (mostly of period two, so called
`blinkers'~\cite{nowakmay}). But, as seen in Fig.~\ref{fig:cd}, when a
regular underlying network is replaced by empirical social networks we
get a complex behavior of both upward and downward spikes and steps
between a number of quasi-equilibrium levels. Other observations
of quasi-stable states in games have been obtained for more complex
update rules and larger strategy
space~\cite{nowak:wsls,brauchli,egui:evo}, or for some special types
of directed networks~\cite{our:pd}.

The steps between quasi-equilibria in Fig.~\ref{fig:cd} suggest a
scenario of relatively few stable states, where mutations on
important vertices may cause a shift from one quasi-stable state to
another. To get a better picture we plot a
histogram over $\rho$ (Fig.~\ref{fig:ph}) that indeed has distinct
peaks. For a given $b$ the position and relative height of the peaks
are $p_m$-independent (if $p_m$ increases, the noise level of the
histograms gets higher) and thus forms a kind of fingerprint of the
network. We note that the histograms obtained from one long run or many
shorter runs (with different random seeds) are the same, this
indicates that the system is self-averaging.

\section{Characteristics of the quasi-stable states}

\begin{figure}
  \centering{\resizebox*{7.2cm}{!}{\includegraphics{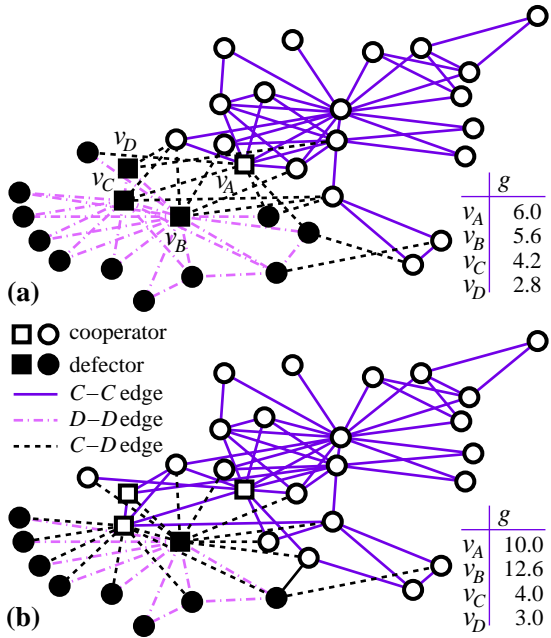}}}
  \caption{The configurations of the intermediate quasi-stable state
    of the karate club network from time steps $779{\,}597$ and
  $779{\,}598$ of the run in Fig.~\ref{fig:cd}(d). The key-players
  mentioned in the discussion are marked as squares.
  }
  \label{fig:kc}
\end{figure}

\begin{figure}
  \centering{\resizebox*{8cm}{!}{\includegraphics{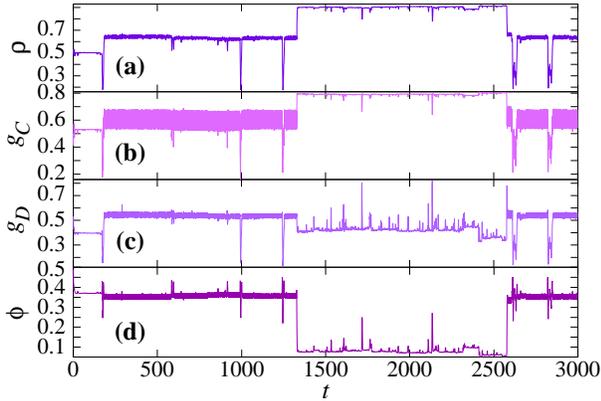}}}
  \caption{(a) cooperator density $\rho$. (b) average cooperator gain
    $g_C$. (c) average defector gain $g_D$. All curves are from the
    same run for the pussokram.com network and parameter values $p_m=
    0.001$, $b=1.35$.
  }
  \label{fig:st}
\end{figure}

Now we turn to the structure of the quasi-stable states. To start with
a concrete example, in Fig.~\ref{fig:kc} we study the karate-club
network at the intermediate $\rho$-level of Fig.~\ref{fig:cd}(d). In the
Fig.~\ref{fig:kc}(a)-configuration~\footnote{We speak of a \emph{configuration} as the set of
  strategies of the players at a given time, and a \emph{state} as the
  set of configurations comprising a period (if the system would evolve
  without mutations).} $v_A$ and $v_B$
scores highest among the border players (players with a neighbor of
the opposite strategy). This makes many players, including the highly
connected $v_C$, change strategy to $C$. In the
Fig.~\ref{fig:kc}(b)-configuration, the defector $v_B$ has enough
cooperators in their neighborhood to score higher than $v_A$, and thus
complete the cycle. We note that a crucial point in sustaining the
stability is that $v_A$ and $v_B$ is not connected. We also note that
if $v_D$ is mutated in the Fig.~\ref{fig:kc}(a) configuration the
system would loose its periodicity and stay in a constant
configuration (until the next mutation). This suggests that the
quasi-equilibria in Fig.~\ref{fig:cd} does not correspond only to one
state, but a set of states that are all close in Hamming distance.

How does the high-$\rho$ states differ from the low-$\rho$ states?
From the update rule to follow the player with locally highest gain,
one expects that defectors gain comparatively much in low-$\rho$ and
vice versa. This is indeed true as shown in Fig.~\ref{fig:st} where we
plot the average defector gain $g_D$ and cooperator gain $g_C$ along
with $\rho$. One may therefore talk about a high-$\rho$ ``cooperator
controlled'' and a low-$\rho$ ``defector controlled'' state. The
average defector gain is directly related to the fraction of boundary
($C$-$D$) edges $\phi$---the total defector gain is the sum of all
boundary edges times $b$, which gives the following expression for
$\phi$:
\begin{equation}
  \phi=\frac{g_D(1-\rho)N}{b M}~.\label{eq}
\end{equation}
The values of $g_D$ (seen in Fig.~\ref{fig:st}(c)) means that
$\sim{\,}40\%$ of the edges are border between cooperators and
defectors in the ``defector controlled'' state. This rather many
considering that the biggest cooperator cluster consists of $\sim
95\%$ of all cooperators (the same figure for defectors is $\sim
63\%$). These observations suggests numerous situations where many
defectors exploit high-degree cooperators, like $v_C$ is being
exploited in Fig.~\ref{fig:kc}(b). (This picture is confirmed under
general circumstances~\cite{our:forthcoming}.)

\begin{figure}
  \centering{\resizebox*{8.2cm}{!}{\includegraphics{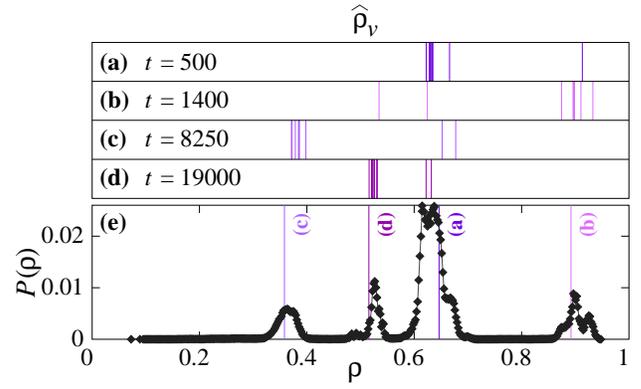}}}
  \caption{
    The impact of one-player mutations. (a)-(d) shows the ten
  $\hat{\rho}_v(t)$ (the eventual cooperator level if $v$, and no
  other vertex, is mutated and the system relaxed) that differs most
  from $\rho(t)$ at the indicated time steps. The original $\rho(t)$
  is indicated together with the $\rho$-histogram in
  (e).
  }
  \label{fig:pi}
\end{figure}

\section{The steps in the cooperator level}

To investigate the mechanism behind the steps of $\rho(t)$
we start by again considering the example of Fig.~\ref{fig:kc}. If the 
most connected vertex $v_B$ (with degree 17) is mutated from $D$ to
$C$ in the configuration of Fig.~\ref{fig:kc}(a) then $g(v_B)=4$ while
$g(v_C)=5.6$. In the next time step $v_B$ would shift to $v_C$'s strategy
$D$ while $v_C$ follows the cooperator $v_A$, so in fact all vertices
except one degree-2 vertex will have the same strategies as without a
mutation on $v_B$. If on the other hand $v_B$ was mutated in the
Fig.~\ref{fig:kc}(b) configuration, all players would cooperate in the
succeeding time step. By this example we conclude that a vertex'
importance with respect to mutation is not only dependent on its
degree but also on the current configuration.

For a little more quantitative approach to this problem we study the
result of a one-player mutation as follows: Starting at time $t$ we
run the system without mutations until one period is completed---this
is to start from a specific configuration $c$ of the local attractor
state. Then we mutate vertex  $v$ and let the system
evolve (without
mutations) until one period is completed. Finally we let the system
complete yet a period and look for the configuration $\hat{c}$ with
the minimal Hamming distance to $c$ over the period (this is to make
the difference zero if the system would relax back to the same state
as prior to the mutation), and set $\hat{\rho}_v=\rho(\hat{c})$. In
Fig.~\ref{fig:pi}(a)-(d) we plot the ten $\hat{\rho}_v$ that differs
most from $\rho(c)$ at four different time steps in four different
quasi-stable states. We observe that the system can change
quasi-stable state due to mutations on single vertices, but at the
same time not go to \emph{any} other quasi-stable state. Furthermore,
there is only a small number of mutations that actually causes a
transition. Just as in the karate club example above the
configurations comprising a state may vary much from one
time-step to another. This fact explains the apparent lack of
transitions to the $\rho\approx 0.35$ state---in fact, the transition
to this state (around $t=8{\,}250$) occurs from a single-player
mutation from a configuration in the $\rho\approx 0.65$ peak. As
expected, all vertices contributing to the lines in
Fig.~\ref{fig:pi}(a)-(d) have high degree (the lowest degree
is $111$). But, as mentioned above, it is not true that a high degree
implies a high importance---for example, of the ten most connected
vertices, on average $5.5$ contributes to the ten most different
$\hat{\rho}_v$ values plotted in Fig.~\ref{fig:pi}(a)-(d). It is thus
not only the structure but the interplay between the structure and
dynamics that gives rise to the quasi-stable equilibria. As a last
point we note that the quasi-equilibria are stable in the absence
of mutations---if the system at the time steps of
Fig.~\ref{fig:pi}(a)-(d) is evolved without mutation, the largest
shift in $\rho$ is $0.008$.

\begin{figure}
  \centering{\resizebox*{8cm}{!}{\includegraphics{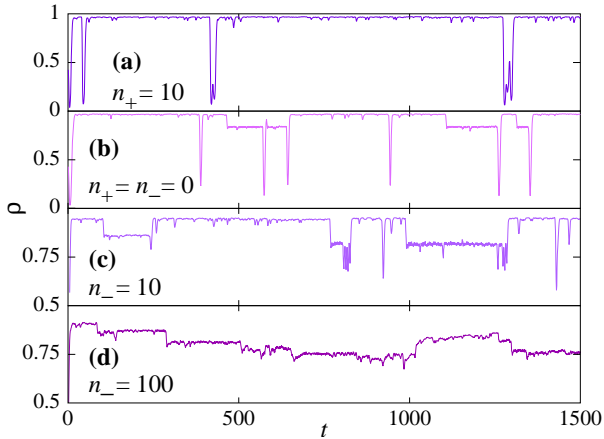}}}
  \caption{ Evolution of $\rho$ for model networks. The network
    parameters are $m=m_0=3$, $N=20{\,}000$. The dynamical parameters
  are $b=1.4$ and $p_m=0.001$.
  }
  \label{fig:ba}
\end{figure}

\section{Minimal structure causing steps and spikes}

What features of the network structure are causing the nonstationary
evolution of the cooperator density? As we have seen, the vertices
have to have a highly skewed degree distribution---so that mutation
on one vertex can cause many vertices, directly or indirectly, change
strategy. Furthermore it should not be the case
that the vertices of highest degree all are interconnected---then
these would all follow the same strategy, and most of the other
vertices too, so nothing more than small fluctuations of $\rho$, or
steps involving almost all vertices, could occur. To test that there
is a bias among the most connected vertices not to attach to each
other, we sample different networks with the same degree sequence by
randomly rewiring the original network with the restriction that the
degree of each vertex is preserved~\cite{roberts:mcmc,maslov:pro}. The
average degree of the maximal subgraph of the ten vertices of highest
degree, $k_{10}$, and the expectation value of the same quantity for
random graphs with the same degree sequence as $G$, $\bar{k}_{10}$,
for the four large networks are shown in Table~\ref{tab:nwk}. (The
value ten is chosen arbitrarily---any number of the same order would
do.) We see that $k_{10}$ is indeed significantly lower than
$\bar{k}_{10}$ for all four networks.  Online interaction represent a
somewhat special type of acquaintance network, even if such networks
are interesting in their own right~\cite{well:comp}, the question if
this structure  exists in `offline' acquaintance networks is open (and
will be discussed further in Ref.~\cite{our:forthcoming}). We also
note that if one runs the PD dynamics on a rewired network one sees
spikes but no quasi-stable states~\cite{our:forthcoming}.

Networks with a broad degree distribution can be generated by, for
example, Barab\'{a}si and Albert's (BA) scale-free network
model~\cite{ba:model}: Start with $m_0$ disconnected vertices, and grow
the network by adding a vertex $v$ with degree $m$ per time step. An
edge from the new vertex are added to an old vertex $w$ with a
probability proportional to $w$'s degree.

We try to model the behavior of the cooperator density by starting
from BA model networks with $m=m_0=3$ and $20{\,}000$ vertices. To
tune how much mutually connected the vertices of highest
degree are, we make the subset of the $n_+$ (or $n_-$) vertices of
highest degree complete by adding the missing edges (or completely
disconnected by removing existing edges). As seen in
Fig.~\ref{fig:ba} the steps and spikes of the real-world networks can
be qualitatively reproduced. The resemblance is closest for removed
edges among a small number of hubs Fig.~\ref{fig:ba}(c). If the edges
are removed within a larger set of hubs, the steps are increasing in
frequency and number until the the structure is completely blurred
out. If all missing edges are added within the top $n_+$
connected vertices, the steps vanishes but the spikes increase in size
and width. An explanation is that when the vertices of highest degree
are fully connected, a mutation of one of these would momentarily
increase the number of cooperators by a large amount and also decrease
the speed of restoration of cooperation. At the same time it is
well-known that cooperation is promoted in highly-connected
regions~\cite{axe:socstr} which explains the high $\rho$-value of the
quasi-stable state.

\section{Summary and conclusions}
In summary, we have studied Nowak and May's spatial PD game on
empirical social networks. We find a complex time-evolution of the
cooperator level characterized by large spikes and transitions between
a number of quasi-stable states. The quasi-stable
states of low $\rho$ are characterized by comparatively large gain for
the defectors and large fraction of boundary edges, in quasi-stable
states with high $\rho$ this situation is reversed. We find that
mutations on high-degree players are likely to be responsible for
transitions between the quasi-stable states; but the outcome of a
mutation of a high-degree vertex is also much dependent on the current
configuration. These findings are presented for some specific sets of
parameter values, but they are qualitatively the same for a broad
range of values. We argue that the structure causing this behavior is
a inhomogeneous degree distribution, and that the number of edges
within the subset highest-degree vertices is relatively low. Based on
this observation we also construct model networks that reproduces the
steps and spikes of the real-world networks. To epitomize, we believe
that we have illustrated how underlying acquaintance patterns can give
rise to a complex time-development of social instability---a picture
that could be empirically testable by carefully arranged social
observations.

\section*{Acknowledgements}
The authors are grateful for comments from C.~Edling, F.~Liljeros and
K.~Sneppen; and acknowledges support from Swedish
Research Council through Contract No.\ 2002-4135 and (B.J.K.) the
Korea Science and Engineering Foundation through Grant No.\
R14-2002-062-01000-0.

\end{document}